# A Be-type star with a  black hole companion

J. Casares[1,2], I. Negueruela[3], M. Ribó[4], I. Ribas[5], J. M. Paredes[4], A. Herrero[1,2] &

S. Simón-Díaz[1,2]

[1] Instituto de Astrofísica de Canarias, E-38205 La Laguna, Santa Cruz de Tenerife, Spain

[2] Departamento de Astrofísica, Universidad de La Laguna, E-38206 La Laguna, Santa Cruz de Tenerife, Spain

[3] Departamento de Física, Ingeniería de Sistemas y Teoría de la Señal, Universidad de Alicante, Apartado. 99, E-03080 Alicante, Spain

[4] Departament d'Astronomia i Meteorologia, Institut de Ciències del Cosmos, Universitat de Barcelona, IEEC-UB, Martí i Franquès 1,  E-08028 Barcelona, Spain

[5] Institut de Ciències de l'Espai – (IEEC-CSIC), Campus UAB, Facultat de Ciències, Torre C5 - parell - 2a planta, E-08193 Bellaterra, Spain

**Stellar-mass black holes have all been discovered through X-ray emission, which arises from the accretion of gas from their binary companions (this gas is either stripped from low-mass stars or supplied as winds from massive ones). Binary evolution models also predict the existence of black holes accreting from the equatorial envelope of rapidly spinning Be-type stars[1-3] (stars of the Be type are hot blue irregular variables showing characteristic spectral emission lines of hydrogen). Of the ~80 Be X-ray binaries known in the Galaxy, however, only pulsating neutron stars have been found as companions[2-4]. A black hole was formally allowed as a solution for the companion to the Be star MWC 656 (ref. 5; also known as HD 215227), although that was based on a single radial velocity curve of the Be star, a mistaken spectral classification[6] and rough estimates of the inclination angle. Here we report observations of an accretion disk line mirroring the orbit of the Be star. This, together with an improved radial velocity curve of the Be star through fitting sharp Fe II profiles from the equatorial disk, and a refined Be classification (to that of a B1.5–B2 III star), reveals a black hole of 3.8 to 6.9 solar masses orbiting MWC 656, the candidate counterpart of the γ-ray source AGL J2241+4454 (refs 5,6). The black hole is X-ray quiescent and fed by a radiatively inefficient accretion flow giving a luminosity less than $1.6 \times 10^{-7}$ times the Eddington luminosity. This implies that Be binaries with black-hole companions are difficult to detect by conventional X-ray surveys.**

The majority of Be X-ray binaries[7] (BeXBs) contain proven neutron stars and are characterized by their transient changes in X-ray luminosity, when episodes of increased accretion onto the compact star are modulated either by the periastron passage or the tidal disruption of the Be circumstellar disk[8]. When in quiescence, they have very low (or even undetectable) X-ray emission. It has been proposed that BeXBs with black holes (Be-BH) are difficult to find because of efficient disk truncation, leading to very long quiescent states[2]. Alternatively, their absence could be driven by binary evolution, with Be-BH binaries having a lower probability of being formed and surviving a common-envelope phase[3].

MWC 656 is a Be star located within the error box of the point-like γ–ray source AGL J2241+4454 (ref. 9). A photometric modulation of 60.37 ± 0.04 days was reported, suggesting that MWC 656 is a member of a binary[6]; this was subsequently confirmed through radial velocities of He I lines from the photosphere of the Be star[5]. The radial velocity curve, though, displays considerable scatter, likely to be caused by filled-in emission from the circumstellar wind contaminating the broad absorption profiles (a common limitation in the analysis of BeXBs: see, for example, ref. 10) and only a tentative orbital solution is available[5]. Here we revisit the 32 Liverpool telescope spectra previously reported[5]. These are complemented with 4 additional Liverpool telescope spectra and a further high-resolution echelle spectrum obtained with the 1.2-m Mercator telescope (see Methods and Extended Data Table 1).

A close-up of the Mercator telescope spectrum is presented in Fig. 1, showing classic Fe II emission lines from the Be circumstellar disk. In addition, a He II 4,686 Å emission line (which was overlooked in a previous work[5]) stands out clearly: its presence is remarkable because it requires temperatures hotter than can be achieved in disks around B-type stars. Further, the He II profile is double-peaked, which is the signature of gas orbiting in a Keplerian geometry[11]. Gaussian fits to the He II profiles in the Liverpool telescope spectra reveal that the centroid of the line is modulated with the 60.37-day orbital period, reaching maximum velocity at photometric phase 0.06 (see Methods and Extended Data Fig. 1). This is approximately in antiphase with the radial velocity curve of the Be star[5], a strong indication that the He II emission arises from gas in an accretion disk around the invisible companion and not from the Be disk. We can therefore use its radial velocity curve to trace the orbit of the Be companion. An eccentric orbital fit to the He II velocities was performed using the Spectroscopic Binary Orbit Program (SBOP[12]), fixing the period to 60.37 days (Methods); the resulting orbital elements are given in Extended Data Table 2. The orbital evolution of the He II line is presented in Fig. 2. The line flux is also found to be modulated with the orbital period (Methods and Extended Data Fig. 1), owing to the presence of an S-wave component swinging between the double peak (see Fig. 2).

In order to improve on the radial velocity curve of the Be star previously reported[5], we fitted the sharp double-peaked profile of the Fe II 4,583 Å emission line with a two-Gaussian model (Methods). Fe II lines are known to arise from the innermost regions of the circumstellar disk[13,14], and therefore reflect the motion of the Be star much more accurately than the broad He I absorptions. The Fe II velocities were also modelled with SBOP, fixing the period to 60.37 days. The eccentric orbital fit results in orbital elements that are consistent with the Fe II orbit being the reflex of the He II orbit, as expected from the motion of two components in a binary system (see Methods and Extended Data Table 2). Only the eccentricities are slightly discrepant, but just at the $1.6\sigma$ level.

Consequently, we modelled the ensemble of Fe II and He II radial velocities with a double-line eccentric binary orbit in SBOP. Figure 3 presents the radial velocity curves of the two emission lines with the best combined solution superimposed. The resulting orbital elements are listed in Table 1. Our solution yields a mass ratio $q = M_2/M_1 = 0.41 \pm 0.07$, which implies a rather massive companion star. A precise determination of the companion's mass requires an accurate spectral classification of the Be star. Accordingly, we have compared the Mercator telescope spectrum with a collection of observed B-type templates, broadened by 330 km s$^{-1}$ to mimic the large rotation velocity in MWC 656 (Methods and Extended Data Fig. 3). Using excitation temperature diagnostics based on several absorption line ratios, we determine a spectral type of B1.5–B2, and the strength of the metallic lines combined with the moderate width of the Balmer absorption wings implies luminosity class III (Methods). Our adopted B1.5–B2 III classification implies a mass of 10–16 solar masses (M$_\odot$) for the Be star (Methods), and hence a companion star of 3.8–6.9 M$_\odot$.

The large dynamical mass of the companion to the Be star MWC 656 is puzzling. A normal main sequence star with such mass would have a spectral type in the range B3–B9 and its spectrum would be easily detected in the optical range. Nor can it be a subdwarf, because these typically have masses in the range 0.8–1.3 M$_\odot$ (ref. 15). The stripped He core of a massive progenitor (that is. a Wolf-Rayet star) is also rejected because it possesses strong winds which show up through intense high excitation emission lines, not present in our spectra. In addition, this should be detected as an ultraviolet excess in the spectral energy distribution, which is not observed in the fluxes available in the literature (Methods). On the other hand, the evidence of a He II accretion disk encircling the companion star strongly points towards the presence of a compact object. The large dynamical mass rules out a white dwarf or a neutron star, so the only viable alternative is a black hole. It should be noted that none of the ~170 BeXBs curently known[4] shows any evidence for an accretion disk, providing circumstantial evidence for a difference in the nature of the compact stars. The accretion disk in MWC 656 is expected to also radiate Balmer and He I lines but these are blended with the corresponding (stronger) emission lines from the Be disk and thus are not detected.

MWC 656 is a key system in the study of BeXBs and massive binary evolution. At a distance $d = 2.6 \pm 0.6$ kpc (Methods) it is relatively nearby and also one of the visually brightest Be binaries[7]. It seems thus reasonable to assume that many other Be-BHs exist in the Galaxy but remain hidden by the lack of transient X-ray activity. Analysis of archival ROSAT images yields an upper limit to the X-ray flux at energies of 0.1–2.4 keV of $1.2 \times 10^{-13}$ erg cm$^{-2}$ s$^{-1}$ (Methods) which, for our estimated distance, translates into an X-ray luminosity $L_x < 1.0 \times 10^{32}$ erg s$^{-1}$ or $< 1.6 \times 10^{-7}$ times de Eddington luminosity, $L_{Edd}$. Therefore, accretion is highly inefficient in MWC 656, akin to accretion on black holes in quiescent low-mass X-ray binaries[16], where accretion disks are truncated at ~$10^2$–$10^4$ Schwarzschild radii and then behave as an advection dominated accretion flow[17].

In the context of disk instability theory, the very low mass-transfer rates expected for BeXBs (with peak values of ~$10^{-11}$ M$_\odot$ yr$^{-1}$ near periastron) lead to extremely long outburst recurrence periods or even to completely suppressed transient activity[18]. It is the dormant condition of the accretion disk together the absence of a solid surface reradiating the accretion energy that makes Be-BHs very difficult to detect through X-ray surveys, thus providing an explanation for the missing Be-BH population. This is in stark contrast with the other black-hole high-mass X-ray binary known in the Galaxy, Cygnus X-1, where an X-ray persistent accretion disk is fed by the powerful wind (~ $10^{-8}$ M$_\odot$ yr$^{-1}$) of an O supergiant star[19].

The detection of a Be-BH is also important to our understanding of BeXB evolution. Whereas the total number of neutron-star BeXBs in the Galaxy depends strongly on the distribution of kick velocities, the number of Be-BHs is very sensitive to the survival probability during the common envelope phase[3]. Modern population synthesis models predict a Galactic number ratio of neutron-star to black-hole BeXBs of 54, for the case of no common envelope survival during the Hertzsprung gap and a Maxwellian distribution of kick velocities with reduced root mean square $\sigma$ =133 km s$^{-1}$ (model C in ref. 3). There are currently ~81 BeXBs known in the Galaxy with ~48 pulsating neutron stars[4,20], and thus our discovery of a black-hole companion to MWC 656 is consistent with these model predictions. However, it should be noted that the X-ray spectra of the remaining BeXBs, whenever they are available, also indicate the presence of a neutron star. Further, in stark contrast with the known BeXBs, MWC 656 has been identified through a claimed γ-ray flare (see Methods) and not by its X-ray activity. This seems to imply that the discovery of Be-BHs is observationally biased, in which case common envelope mergers would be less frequent than commonly assumed and/or neutron star kicks would be best described by the radio pulsar birth velocity distribution[3]. Last, it is interesting to note that MWC 656 will probably evolve into a black-hole/neutron-star binary, a potential source of strong gravitational waves and a short γ–ray burst (Methods).

**METHODS SUMMARY**

The Fe II line was fitted with a two-Gaussian model with Gaussian positions, intensities and separation left as free parameters. The Fe II velocities were obtained from the mean of the Gaussians offset with respect to the rest velocity at 4,583.837 Å. A detailed radial velocity study of the He II profile was performed using the double-Gaussian technique[21], and this shows that the systemic velocity is pushed down in the line core by the S-wave component while the phasing and velocity semiamplitude remain very stable. Therefore, a +30 km s$^{-1}$ offset was applied to the He II velocities before fitting all the data points with a double-line orbital model. The S-wave also modulates the He II flux with the orbital period, and its phasing can be interpreted as either enhanced mass transfer during periastron or the visibility of a hotspot in the outer accretion disk. The spectral classification of the star was obtained by direct comparison with a range of templates conveniently broadened, and the best match is provided by the B1.5 III star HD 214993. We used several calibrations available in the literature to constrain the mass of MWC 656 from its spectral type, including evolutionary tracks, a dynamical determinations from the detached eclipsing binary V380 Cyg and robust lower limits from dynamical masses of main sequence stars. The distance is obtained through combining the absolute magnitude of B1.5-2 III stars with the observed brightness of MWC 656, corrected for intestellar reddening. Upper limits to the X-ray flux of MWC 656 are derived from archival ROSAT and Swift pointings, using a neutral hydrogen column density $N_{\mathrm{H}} = 1.4 \times 10^{21}$ cm$^{-2}$ and a photon index $\Gamma = 2.0$, typical of black holes in quiescence. We futher discuss the future evolution of MWC 656 and its fate, a possible black-hole/neutron-star binary.

**References**

1 Raguzova, N.V. & Lipunov, V.M. The evolutionary evidence for Be/black hole binaries *Astron. Astrophys.* **349**, 505-510 (1999).

2 Zhang, F., Li, X.-D. & Wang, Z.-R., Where are the Be/Black Hole binaries? *Astrophys. J.* **603**, 663-668 (2004).

3 Belczynski, K. & Ziólkowski, J., On the apparent lack of Be X-Ray binaries with black holes. *Astrophys. J.* **707**, 870-877 (2009).

4 Ziólkowski, J. & Belczynski, K., On the apparent lack of Be X-Ray binaries with black holes in the Galaxy and in the Magellanic Clouds. *IAU Symposium* **275,** 329-330 (2011) arXiv:1111.2330.

5 Casares, J. *et al.,* On the binary nature of the γ-ray sources AGL J2241+4454 (= MWC 656) and HESS J0632+057 (= MWC 148). *Mon. Not. R. Astron. Soc.* **421,** 1103-1112 (2012).

6 Williams, S. J. *et al.*, The Be Star HD 215227: a candidate gamma-ray binary. *Astroph. J.* **723**, L93-L97 (2010).

7 Reig, P., Be/X-ray binaries. *Astrophys. Space Sci.* **332**, 1-29 (2011).

8 Okazaki, A. T. & Negueruela, I., A natural explanation for periodic X-ray outbursts in Be/X-ray binaries. *Astron. Astrophys*. **377,** 161-174 (2001).

9 Lucarelli, F. *et al*., AGILE detection of the new unidentified gamma-ray source AGL J2241+4454. *Astron. Teleg.* **2761,** 1 (2010).

10 Ballereau, D., Chauville, J. & Zorec, J., High-resolution spectroscopy of southern and equatorial Be stars: flux excess at λ4471 Å. *Astron. Astrophys. Suppl. Ser.* **111**, 423-455 (1995).

11 Smak, J., On the rotational velocities of gaseous rings in close binary systems. *Acta. Astronomica* **19,** 155-164 (1969).

12 Etzel, P., SBOP: Spectroscopic Binary Orbit Program. (San Diego State Univ., 2004).

13 Hanuschik, R. W., On the structure of Be star disks. *Astron. Astrophys.* **308**, 170-179 (1996).

14 Arias, M. L. *et al*., Fe II emission lines in Be stars. I. Empirical diagnostic of physical conditions in the circumstellar discs. *Astron. Astrophys.* **460**, 821-829 (2006).

15 Peters, G. J., Pewett, T. D., Gies, D. R., Touhami, Y. N. & Grundstrom, E. D., Far-ultraviolet detection of the suspected subdwarf companion to the Be star 59 Cygni. *Astrophys. J.* **765**, 2-9 (2013).

16 Garcia, M. R. *et al*., New evidence for black hole event horizons from Chandra. *Astrophys. J.* **553**, L47-L50 (2001).

17 Esin, A. A., McClintock, J. E. & Narayan, R., Advection-dominated accretion and the spectral states of black hole X-ray binaries: application to Nova Muscae 1991. *Astrophys. J.* **489,** 865-889 (1997).

18 Menou, K., Narayan, R. & Lasota, J.-P., A population of faint nontransient low-mass black hole binaries. *Astrophys. J.* **513**, 811-826 (1999).

19 Coriat, M., Fender, R. P. & Dubus, G., Revisiting a fundamental test of the disc instability model for X-ray binaries. *Mon. Not. R. Astron. Soc.* **424**, 1991-2001 (2012).

20 Linden, T., Valsecchi, F. & Kalogera, V., On the rarity of X-ray binaries with naked helium donors. *Astrophys. J.* **748,** 114-121 (2012).

21 Schneider, D.P. & Young, P., The magnetic maw of 2A 0311-227. *Astrophys. J.* **238**, 946-954 (1980).

22 Smak, J., On the S-wave components of the emission lines in the spectra of cataclysmic variables. *Acta. Astron*omica **35**, 351-367 (1985).

**Acknowledgments.** We thank T. Maccarone and P. Charles for comments on the paper. This work made use of the molly software package developed by T. R. Marsh. The Liverpool telescope and the Mercator telescope are operated on the island of La Palma by the Liverpool John Moores University and the University of Leuven/Observatory of Geneva, respectively, in the Spanish Observatorio del Roque de los Muchachos of the Instituto de Astrofisica de Canarias. The Liverpool telescope is funded by the UK Science and Technology Facilities Council. This research was supported by the Spanish MINECO and FEDER under grants AYA2010-18080, AYA2010-21782-C03-01, AYA2010-21967-C05-04/05, AYA2012-39364-C02-01/02, AYA2012-39612-C03-01, FPA2010-22056-C06-02 and SEV2011-0187-01; it was also funded by the grant PID 2010119 from the Gobierno de Canarias. J.M.P. acknowledges financial support from ICREA Academia.

**Authors Contribution** J.C. performed the radial velociy analysis of the spectra and wrote the paper. I.N. obtained the Mercator spectrum and contributed to the interpretation of the data. I.R. computed the eccentric orbital fits to the radial velocity curves. M.R. calculated the distance and X-ray luminosity, and contributed to the interpretation of the data. J.M.P. also contributed to the interpretation of the data. A.H. computed the rotational broadening of the star and, together with I.N., performed the spectral calibration of the star. S.S-D. observed the standard stars and reduced the Mercator spectra. J.M.P. and M.R. assisted in writing the section on γ-ray binaries in Methods.

**Table 1: Orbital elements for MWC 656.**

| Parameter | MWC 656 |
|---|---|
| $P_{orb}$ (days) | 60.37 (fixed) |
| $T_0$ (HJD-2,450,000) | 3243.70±4.30 |
| $e$ | 0.10±0.04 |
| $\omega$ (deg) | 163.0±25.6 |
| $\gamma$ (km s$^{-1}$) | −14.1±2.1 |
| $K_1$ (km s$^{-1}$) | 32.0±5.3 |
| $K_2$ (km s$^{-1}$) | 78.1±3.2 |
| $a_1 \sin i$ (R$_\odot$) | 38.0±6.3 |
| $a_2 \sin i$ (R$_\odot$) | 92.8±3.8 |
| $M_1 \sin^3 i$ (M$_\odot$) | 5.83±0.70 |
| $M_2 \sin^3 i$ (M$_\odot$) | 2.39±0.48 |
| $M_2/M_1$ | 0.41±0.07 |
| $\sigma_f$ (km s$^{-1}$) | 16.7 |

The solution was obtained from a combined fit to the radial velocity curves of the He II 4,686 Å and Fe II 4,583 Å lines. The orbital period $P_{orb}$ has been fixed to the photometric value[6]. $T_0$ is the epoch of periastron (where HJD refers to Heliocentric Julian Date), $e$ the orbit eccentricity, $\omega$ the longitude of periastron, $\gamma$ the systemic velocity, $K$ the velocity semiamplitude, $a$ the semimajor axis, $i$ the binary inclination, $M$ the stellar mass and $\sigma_f$ the rms of the fit. Subscripts 1 and 2 refer to the primary (Be star) and secondary (companion) components, respectively. $T_0$ implies that periastron passage occurs at photometric phase 0.01 ± 0.10.

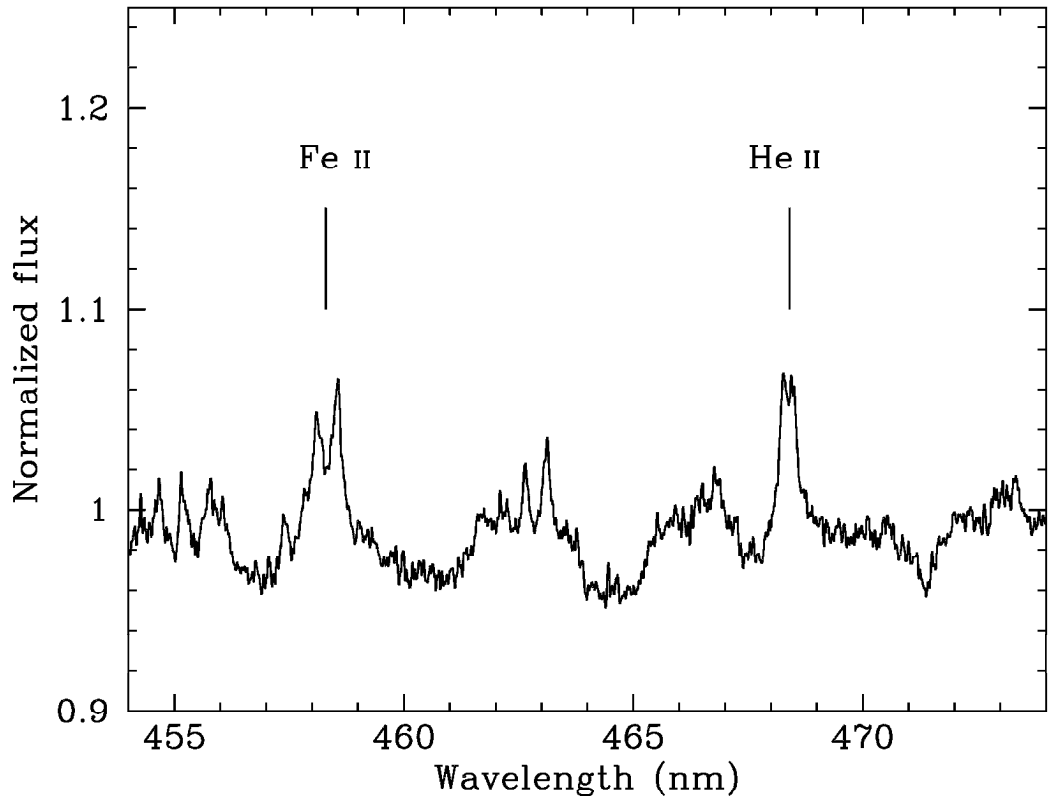

**Figure 1: Optical spectrum of MWC 656, obtained with the Mercator telescope.** The spectrum has been rebinned to 0.1 Å per pixel and smoothed through a 3-pixel Gaussian bandpass. The emission lines Fe II 4583 Å (of multiplet 38) and He II 4,686 Å are indicated. Whereas Fe II is formed in the Be circumstellar disk, He II arises from gas encircling the companion star. Several other circumstellar Fe II lines are detected in the spectrum (for example, 4,549, 4,555, 4,629, 4,666, 4,731 Å) but they are weaker and/or severely blended with other emission lines and photospheric absorptions.

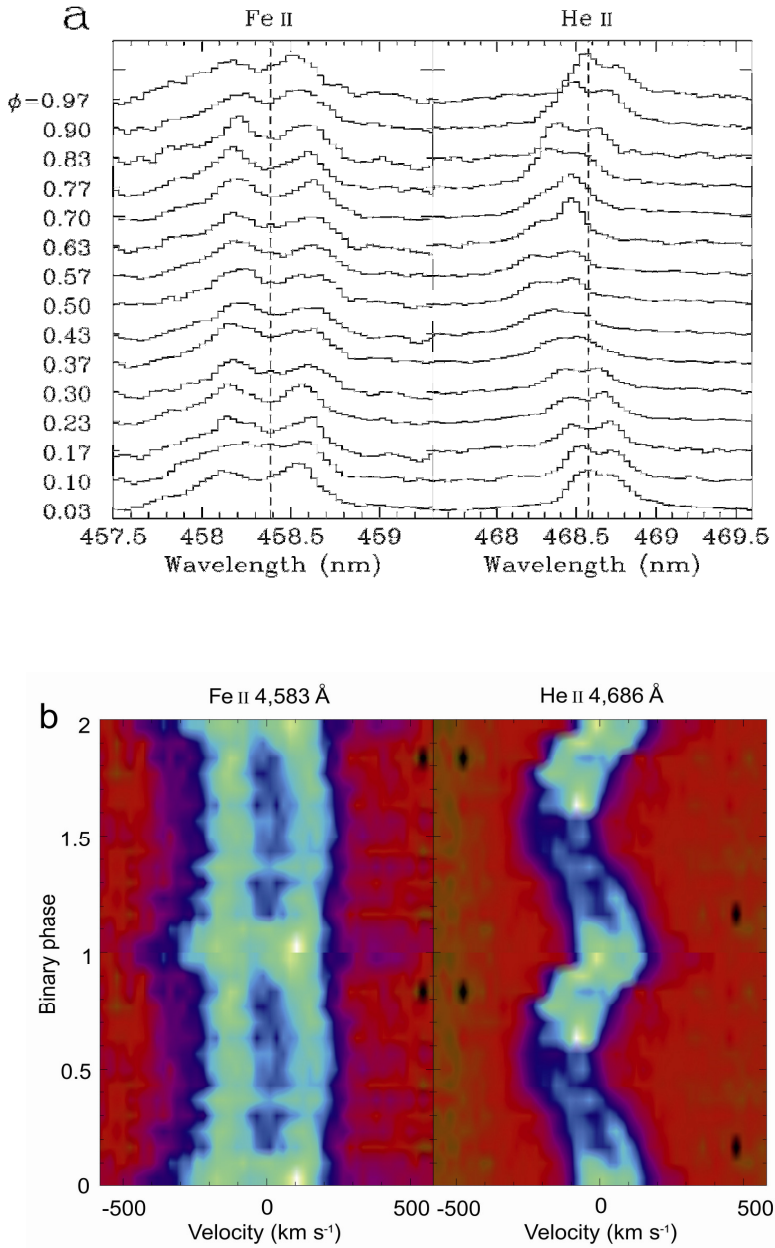

**Figure 2: Orbital evolution of the Fe II 4,583 Å and He II 4,686 Å emission lines. a,** Sequence of spectra folded into 15 phase bins using the photometric ephemeris[6]. $\Phi$ indicates the binary orbital phase. Vertical dashed lines indicate the central wavelength of each line. For the sake of clarity, the flux of the Fe II line has been arbitrarily scaled by a factor of 2 with respect to He II. The He II double peak is distorted by an S-wave component, typically associated witg a bright spot or asymmetry in the outer accretion disk[22]. **b,** Trailed intensity image of the two lines constructed from the phase binned spectra. Two orbital cycles are displayed for clarity. The colour scale indicate counts normalized to the continuum, with the black colour corresponding to 0.98 and the white colour to 1.08 in Fe II and 1.16 in He II.

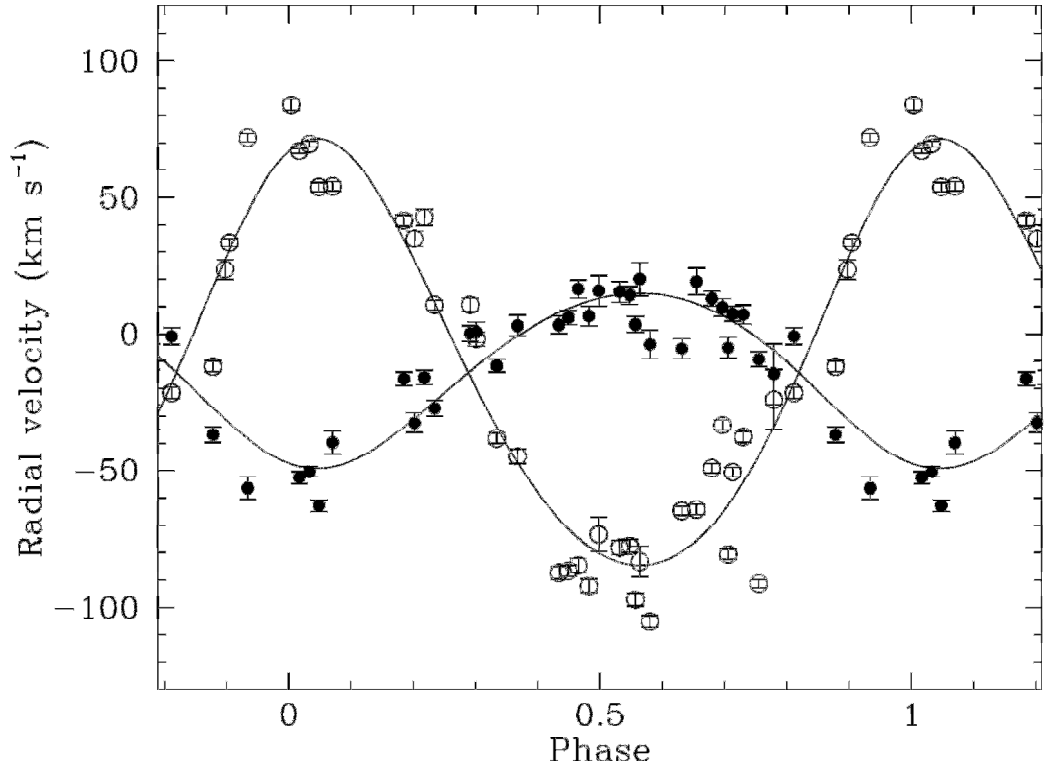

**Figure 3: The radial velocity curves of the Be star and its companion.** Solid circles indicate velocities of the Be star, as obtained from the Fe ɪɪ 4,583 Å line, open circles those of the companion star, extracted from the He ɪɪ 4,686 Å line. Error bars, 1 s.d. The best combined orbital fit is overplotted. The velocities have been folded with the photometric ephemeris[6] as in Fig. 2. A shift of +30 kms$^{-1}$ was applied to the He ɪɪ velocities to compensate for the S-wave contamination (Methods and Extended Data Fig. 2). Two Fe ɪɪ velocities (not shown) were found to deviate by >2 $\sigma_f$ (where $\sigma_f$ is the r.m.s. of the fit) from the best solution and were excluded from the fit, but they have a negligible impact in the final orbital elements.

# METHODS

## Spectroscopic observations

Thirty two 10-min spectra of MWC 656 were obtained between 23 April and 28 July 2011 using the Fibre-fed RObotic Dual-beam Optical Spectrograph (FRODOspec) on the robotic 2.0-m Liverpool telescope at the Observatorio del Roque de Los Muchachos on La Palma (Spain). Full details on these observations are given elsewhere[5]. Four additional FRODOspec spectra were collected on the nights of 28 May and 2–4 June 2012 using identical instrument configuration. We also used the the High Efficiency and Resolution Mercator Echelle Spectrograph (HERMES) on the 1.2-m MERCATOR telescope to obtain a 15-min spectrum on the night of 26 October 2012. We employed the high resolution mode in HERMES which yields a resolving power R=85,000 accross the entire optical range between 3,770–9,000 Å. A series of B-type MK standards (where MK refers to the Morgan-Keenan spectral classification) were also observed with HERMES on the nights of 14 June and 9 November 2011 using identical setup as for MWC 656. The automatic pipeline products were used for the extraction and calibration of the Liverpool telescope and Mercator telescope data. A full log of the observations is presented in Extended Data Table 1.

## Radial velocity analysis

Radial velocities were extracted from the He $\text{\sc ii}$ 4,686 Å emission by fitting a single Gaussian to the line profile in the 36 Liverpool telescope spectra. A least-squares sine fit to the velocity points yields an orbital period of 59.5 ± 0.6 days (Extended Data Fig. 1a) which is consistent within 1.4 $\sigma$ with the (more accurate) photometric period determination[6]. The difference is also explained by the fact that our radial velocities only cover two full orbital cycles. Consequently, we henceforth adopt 60.37 ± 0.04 days as the true orbital period of the binary. The radial velocities were modelled with an eccentric binary orbit using the code SBOP[12], with the period fixed to 60.37 days. Individual points were weighted proportionally to $1/\sigma^2$, where $\sigma$ is the radial velocity uncertainty. We adjusted the following orbital parameters: epoch of periastron ($T_0$), eccentricity ($e$), argument of the periastron ($\omega$), systemic velocity ($\gamma$) and velocity semiamplitude ($K$). The resulting orbital elements are listed in the first column of Extended Data Table 2, together with their implied fundamental binary parameters. The fitted solution yields maximum velocity at phase 0.06, where phase 0 is arbitrarily set to HJD 2453243.3 or the epoch of maximum brightness in the photometric light curve[6]. This is approximately in antiphase with the radial velocity of the Be star previously determined[5], and therefore the He II emission most probably follows the orbit of the companion star.

To further constrain the orbital motion of the Be star we measured radial velocities from the Fe II 4,583 Å emission line by fitting a two-Gaussian model to the individual profiles. The Gaussians are set to have identical width but their intensities and separation are allowed to vary in order to account for possible profile changes. In any case, these three free parameters are found to be very stable, with differences which are always within 10 %. Only in one case was the double-peak separation found to be 13% smaller than the average. This corresponds to one spectrum where the profile appears blurred. For this particular case we decided to fix the Gaussian separation to the average value, 270 km s$^{-1}$. The radial velocity of the Fe II line was then obtained from the center of the double Gaussian model relative to the line rest velocity at 4,583.837 Å. We subsequently fitted the radial velocity points with an eccentric orbit model using SBOP after fixing the orbital period to 60.37 days and weighting data points proportionally to $1/\sigma^2$, as before. The best set of fitting parameters are listed in the second column of Extende Data Table 2.

The two independent solutions presented in Extended Data Table 2 give periastron phases $T_0$ which are consistent at 1.0 $\sigma$. Also, the arguments of periastron are in antiphase within 1.0 $\sigma$ uncertainties, i.e. $\omega_{FeII} = \omega_{HeII} + 180^o$. Further, the eccentricities are consistent at the 1.6$\sigma$ level. These are all strong indications that both radial velocity curves display the reflex motion of two components in a binary system, thus endorsing fitting a combined double orbit to the ensemble of 72 He II and Fe II data points. In any case, the radial velocity amplitudes of the single line orbits are found to be very robust and in excellent agreement with the combined double-lined solution presented in the main text. These are the key parameters constraining the binary mass ratio and thus the nature of the companion to the Be star.

## The diagnostic diagram

Extended Data Table 2 shows a 30 km s$^{-1}$ difference between the systemic velocities of the Fe II and He II solutions, which we interpret as due to contamination by the S-wave component seen in the core of the He II profile[22]. To test this, we measured radial velocities from different velocity bands of the He II line profile, using the double-Gaussian technique[21]. To further increase the signal-to-noise ratio, we averaged the 36 Liverpool telescope spectra into 15 phase bins, relative to the photometric ephemeris. Every phase-binned spectrum was convolved with a two-Gaussian passband where the Gaussian separation was varied between $a$ =200–600 km s$^{-1}$ in steps of 50 km s$^{-1}$. Gaussian separations $a < 200$ km s$^{-1}$ produce radial velocities which are corrupted by the variable shape of double-peaked profile, while for $a > 600$ km s$^{-1}$ the line flux becomes too weak. Each radial velocity curve was subsequently fitted with the expression $V(\varphi)$ $=\gamma+K$ sin $2\pi(\varphi-\varphi_0)$, where $\varphi$ is the binary phase, and the evolution of the fitting parameters $K$, $\gamma$ and $\varphi_0$ versus the Gaussian separation $a$ is displayed in Extended Data Fig. 2 as a diagnostic diagram[23]. Note that we prefer to fit a simple sine wave model rather than a full eccentric orbit because the extra fitting parameters (that is, $e$, $\omega$ and periastron phase) become poorly constrained as we approach the noisy wings of the line profile. Given the small orbital eccentricity the fitted sine wave parameters are still meaningful, although they should be taken as a first approximation to the true values because of the model oversimplification.

The high velocity wings of the He II profile are formed in the inner regions of the accretion disk and are less affected by the core S-wave component. Therefore, they are expected to closely trace the motion of the compact star. Extended Data Fig. 2 shows that, as we move away from the line core, the systemic velocity rises quickly from about −46 km s$^{-1}$ to about −25 km s$^{-1}$, thus approaching the value of the Fe II solution. This demonstrates that the Fe II velocities provide a more accurate description of the true binary systemic velocity than the centroid of the He II line and, therefore, we decide to apply a +30 km s$^{-1}$ offset to the latter. Beyond $a > 500$ km s$^{-1}$ the continuum noise begins to dominate the radial velocities, as indicated by the steeper rise in the control parameter $\sigma(K)/K$ (ref. 23), and thus the fitted parameters are less reliable. The diagram also illustrates that both phasing and velocity semiamplitude are very stable in the interval $a$ =200–500 km s$^{-1}$, ranging between $\varphi_0$ =0.77–0.82 and $K$ =76–92 km s$^{-1}$. Beyond $a >$ 500 km s$^{-1}$ $K$ drops below 70 km s$^{-1}$ and hence there is a possibility that the velocity semiamplitude of the compact star was overestimated in Table 1. Note, however, that this would raise the binary mass ratio which would make even stronger the argument for a black hole companion.

## Orbital modulation of the He II Flux

Extended Data Fig. 1b shows that the equivalent width of the He II 4,686 Å line is strongly modulated with the orbital period. It peaks at phase ~0.9, which is very close to the maximum in the photometric light curve (that is, phase 0 by convention). On the other hand, the amplitude of the equivalent width modulation is an order of magnitude larger (~40 %) than that of the photometric light curve[6,24]. These two arguments imply that the equivalent width variability is driven by true changes in the line flux rather than in the continuum.

Extended Data Fig. 1 also reveals that the maximum equivalent width (phase 0.93 ± 0.04) almost coincides with the peak in the He II radial velocity curve, when the compact star is receding from us at maximum speed. The latter agrees well with the time of maximum visibility of the hotspot or shock region between the gas stream and the accretion disk. Alternatively, the modulation of the He II flux can be interpreted as caused by enhanced mass transfer near periastron, which is constrained to phase 0.01 ± 0.10 by our orbital solution.

## Spectral classification, mass and distance to MWC 656

The spectral classification of MWC 656 is complicated by the effects of fast rotation, and by the presence of many (mostly Fe II) emission lines affecting many of the features useful for this purpose. To provide an improved classification, our high-quality Mercator telescope+HERMES spectrum was compared to the spectra of several MK standards taken with the same instrumentation and set-up as for MWC 656. Before this, we measured the rotational broadening in MWC 656 by applying the Fourier technique[25], combined with a goodness-of-fit method[26], to the He I absorption profiles (4,387, 4,471, 4,922 Å) and the Si III line 4,552 Å. This technique allows to disentangle the different contributions to the line broadening, with the first zeroes of the transformed profile due to rotation. We obtain $v \sin i = 330 \pm 30$ km s$^{-1}$, with the error reflecting the dispersion of the individual lines, in good agreement with the value previously reported[5]. All our MK standards have very small intrinsic broadenings[27], and thus they were subsequently broadened by 330 km s$^{-1}$ in order to reproduce the observed broadening in MWC 656. This was done by convolving the MK spectra with a Gray rotational profile[25], using a limb-darkening coefficient $\varepsilon = 0.34$ which is appropriate for the stellar parameters of our target (see below) and the spectral range of interest.

The narrow wings of the Balmer lines definitely indicate that the star is a giant, while the strength of the O II spectrum makes it B2 or earlier, in contrast with a previous classification[6] which reported a B3 IV. The absence of Si IV and He II absorption lines places MWC 656 in the B1–B2 range. As shown in Extended Data Fig. 3, the best match to the overall spectrum is provided by the B1.5 III standard HD 214993, although MWC 656 has rather stronger N II features. Nitrogen enhancement is frequently found in fast rotators[28,29], and understood as a natural product of stellar evolution[30]. Nitrogen enhancement is often accompanied by C depletion. Therefore, we cannot completely rule out the possibility that MWC 656 is a B2 III giant with some C depletion (compare to the spectrum of the MK standard HD 35468 in Extended Data Fig. 3), but the strength of Si III and O II lines, which are the features most sensitive to temperature in this spectral range, strongly supports a B1.5 III classification.

Armed with the spectral classification we can now set constraints to the mass of the Be star using the several calibrations available in literature. Based on evolutionary tracks[31], giant stars in the range B1–B2 have mass 12–17 $M_\odot$. On the other hand, the most precise calibrations are provided by dynamical determinations in detached eclipsing binaries but unfortunately data on B1–B2 giants is very scarce. The closest example is provided by the B1.5 III star in V380 Cyg, where a dynamical mass of 13.1 ± 0.2 $M_\odot$ has been reported[32]. No further analogies are found in a recent exhaustive compilation of detached binaries[33]. In any case, robust lower limits to the mass of our target are provided by dynamical masses of main sequence stars. For instance ref. 33 give 11 $M_\odot$ for B1 V and 9 $M_\odot$ for B2 V while ref. 34 yields 10 $M_\odot$ for B1.5 V. Taking into account all the above, we support a mass range of 10–16 $M_\odot$ for the Be star in MWC 656. We also remark that the above mass range might be an underestimate because of the large rotational velocity of MWC 656. This causes the star to appear cooler and slightly less luminous than an object of the same stellar mass rotating at lower speed. This effect could increase our estimated mass by ~10–15% which would raise the black-hole mass even more[35].

We can also estimate the distance to MWC 656 by comparing the absolute magnitudes of B1.5–B2 III stars with the observed brightness, corrected for interstellar reddening. To estimate the latter, we have constructed the spectral energy distribution (SED) of MWC 656 in the ultraviolet and optical bands[36,37]. The SED was fitted in the range 150–500 nm (where the contribution of the Be disk is marginal) with TLUSTY[38] and FASTWIND[39] models reddened by several amounts, and the best result is obtained for $E(B - V) = 0.24$. Adopting an absolute visual magnitude $M_V$ in the range from −3.7 to −4.5, based on the calibration of ref. 31, and V = 8.75 ± 0.10 (ref. 40), we derive a distance of 2.6 ± 0.6 kpc. Note that in this calculation we have neglected the contribution to the optical light from both the circumstellar Be disk and the black-hole accretion disk. The fact that MWC 656 shows photometric variability modulated with the orbital period argues for some contribution from non-stellar sources to the optical flux.

However, the modulation has a semiamplitude of only 0.020 mag, suggesting that these other contributions have a very modest effect[6,24] and, therefore, we can safely conclude that the Be star dominates the observed optical flux. In any case, the large error in our distance estimate encompasses the uncertainty introduced by the small contribution of non-stellar light sources to the observed optical flux. Incidentally, the SED does not show any ultraviolet excess as would be expected if the companion to the Be star were a stripped He core. This gives additional support for the presence of a black hole in MWC 656.

## X-ray analysis

Inspection of the HEASARC archives at the position of MWC 656 reveals X-ray observations by ROSAT and Swift. The ROSAT/Position Sensitive Proportional Counters (PSPC) observations (0.1–2.4 keV) were conducted on 7–11 July 1993 (orbital phases 0.63 to 0.70) and consisted on five pointings of ~3–6 ks each. The Swift/X-Ray Telescope (XRT) observations (0.3–10 keV) were conducted on 8 March 2011 (orbital phase 0.52) and consisted on two 1-ks pointing separated around 3 h.

Both data sets have been analyzed using standard procedures within HEASOFT V 6.12. Assuming a photon index $\Gamma = 2.0$ (typical of black holes in quiescence[41]) and a hydrogen column density $N_H = 1.4 \times 10^{21}$ cm$^{-2}$ (using the obtained $E(B-V)$ and the relation with $N_H$ of ref. 42) the following upper limits (at the 90% confidence) are obtained: $F_{0.1–2.4\,keV} < 1.2 \times 10^{-13}$ erg cm$^{-2}$ s$^{-1}$ for ROSAT/PSPC and $F_{0.3–10\,keV} < 4.6 \times 10^{-13}$ erg cm$^{-2}$ s$^{-1}$ for Swift/XRT.

## Gamma-ray association

MWC 656 has been proposed as the optical counterpart of the transient GeV γ-ray source AGL J2241+4454 (refs 5,6). The association is, however, uncertain because AGL J2241+4454 was only detected during a 2-day activity period and it has a position error circle radius of 0.6$^o$ (ref. 9). A few binary systems have been detected at GeV and/or TeV energies[43], all showing orbitally modulated γ-ray emission and most of them thought to contain young non-accreting pulsars. In contrast, the accreting black hole Cygnus X-1 showed evidence of TeV emission during less than one day[44], and also two different short transient episodes of GeV emission[45]. The black-hole nature of MWC 656 and its putative flaring γ-ray activity are reminiscent to those of the well known black hole Cygnus X-1 but, remarkably, the accretion luminosity in MWC 656 is typically lower by more than ~5 orders of magnitude.

## A candidate black-hole/neutron-star progenitor

The future evolution of MWC 656 will probably lead to a black-hole/neutron-star binary[46]. During the red giant phase the 13 M$_\odot$ Be star will expand by several hundred solar radii[47], thus engulfing the black hole. Mass transfer from the expanding Be star onto the black hole will be dynamically unstable and a common envelope will ensue. This is a highly dissipative process which leads to spiral-in of the black hole, efficient circularization of the orbit and the ejection of the Be star envelope. The outcome of the common envelope phase will then be a 2.9 M$_\odot$ He star[48] and the present ~5 M$_\odot$ black hole companion in a close circular orbit. In the event of a symmetric core collapse the new born neutron-star/black-hole binary will remain bound because less than half the total initial mass is expelled in the explosion[49]. In the case of an asymmetric supernova, the binary survival will depend on the magnitude and direction of the kick.

Black-hole/neutron-star binaries, which have not been detected yet, are instrumental in providing fundamental tests of gravitational theories, strong sources of gravitational waves and prime candidates for the production of short γ-ray burst through coalescence[50-52]. The fate of MWC 656 as a possible black-hole/neutron-star binary is very relevant because it provides tight empirical constraints on detection rates for gravitational wave observatories, such as advanced LIGO/VIRGO[53].

### References

23 Shafter, A. W., Szkody, P. & Thorstensen, J. R., X-ray and optical observations of the ultrashort period dwarf nova SW Ursae Majoris - a likely new DQ Herculis star. *Astrophys. J.* **308**, 765-780 (1986).

24 Paredes-Fortuny, X., Ribó, M., Fors, O. & Núñez, J., Optical photometric monitoring of gamma-ray binaries. *Am. Inst. Phys. Conf. Ser.*, **1505**, 390–393. (2012)

25 Gray, D.F., *The Observations and Analysis of Stellar Photospheres* (CUP 20, Wiley-Interscience, New York, 1992).

26 Ryans, R. S. *et al*., Macroturbulent and rotational broadening in the spectra of B-type supergiants *Astron. Astrophys*. **336**, 577-586 (2002).

27 Abt, H.A., Levato, H. & Grosso, M., Rotational velocities of B stars. *Astrophys. J.* **573**, 359-365 (2002).

28 Howarth, I. D. & Smith, K. C., Rotational mixing in early-type main-sequence stars. *Mon. Not. R. Astron. Soc.* **327**, 353-368 (2001).

29 Walborn, N. R. *et al*., Further Results from the Galactic O-Star Spectroscopic Survey: Rapidly Rotating Late ON Giants. *Astron. J.* **142,** 150-156 (2011).

30 Przybilla, N., Firnstein, M., Nieva, M. F., Meynet, G. & Maeder, A., Mixing of CNO-cycled matter in massive stars. *Astron. Astrophys*. **517**, A38-A43 (2010).

31 Straizys, V. & Kuriliene, G., Fundamental stellar parameters derived from the evolutionary tracks. *Astrophys. Space Sci.* **80,** 353-368 (1981).

32 Pavlovski, K. *et al*., Chemical evolution of high-mass stars in close binaries - II. The evolved component of the eclipsing binary V380 Cygni. *Mon. Not. R. Astron. Soc.* **400**, 791-804 (2009).

33 Torres, G., Andersen, J. & Giménez, A., Accurate masses and radii of normal stars: modern results and applications. *Astron. & Astrophys. Rev.* **18**, 67-126 (2010).

34 Harmanec, P., Stellar masses and radii based on modern binary data *Bull. Astron. Inst. Czech.* **39**, 329-345 (1988).

35 Negueruela, I. *et al*., Astrophysical parameters of LS 2883 and implications for the PSR B1259-63 gamma-ray binary. *Astroph. J.* **732**, L11-L15 (2011).

36 Thompson, G.I. *et al*., Catalogue of stellar ultraviolet fluxes. A Compilation of Absolute Stellar Fluxes Measured by the Sky Survey Telescope (S2/68) Aboard the ESRO Satellite TD-1, (Science Research Council, UK, 1978).

37 Merrill, P.W. & Burwell, C.G., Supplement to the Mount Wilson catalogue and bibliography of stars of classes B and A whose spectra have bright hydrogen lines. *Astrophys. J.* **98**, 153-184 (1943).

38 Hubeny, I. & Lanz T., NLTE line blanketed model atmospheres of hot stars. I. Hybrid complete linearization/accelerated lambda iteration method. *Astroph. J. (Letters)* **439**, 875-904 (1995).

39 Puls, J. *et al*., Atmospheric NLTE-models for the spectroscopic analysis of blue stars with winds. II. Line-blanketed models *Astron. Astrophys*. **435**, 669-698 (200).

40 Nicolet, B., Catalogue of homogeneous data in the UBV photoelectric photometric system. *Astron. Astrophys. Suppl. Ser.* **34**, 1-49 (1978).

41 Plotkin, R. M., Gallo, E. & Jonker, P.G., The X-ray spectral evolution of galactic black hole X-ray binaries toward quiescence *Astrophys. J.* **773**, 59-74 (2013).

42 Bohlin, R.C., Savage, B.D. & Drake, J.F., A survey of interstellar H ɪ from L-alpha absorption measurements. II. *Astrophys. J.* **224,** 132-134 (1978).

43 Mirabel, I.F., Gamma-Ray binaries revealed *Science* **335**, 175-176 (2012).

44 Albert, J., *et al*., Very high energy gamma-ray radiation from the stellar mass black hole binary Cygnus X-1 *Astroph. J.* **665**, L51-L54 (2007).

45 Sabatini, S., *et al*., Gamma-ray observations of Cygnus X-1 above 100 MeV in the hard and soft states. *Astrophys. J.* **766**, 83-97 (2013).

46 Lipunov, V.M., Postnov, K.A., Prokhorov, M.E. & Osminkin, E.Yu, Binary radiopulsars with black holes *Astroph. J.* **423**, L121-L124 (1994).

47 Schaller, G., Schaerer, D., Meynet, G. & Maeder, A., New grids of stellar models from 0.8 to 120 solar masses at Z = 0.020 and Z = 0.001 *Astron. Astrophys.* **96** *(Suppl.)* 269-331 (1992).

48 Woosley, S.E. & Weaver, T.A., The evolution and explosion of massive stars. II. Explosive hydrodynamics and nucleosynthesis *Astroph. J. Suppl. Ser.* **101**, 181-235 (1995).

49 Boersma, J., Mathematical theory of the two-body problem with one of the masses decreasing with time *Bull. Astron. Inst. Neth.* **15**, 291-301 (1961).

50 Narayan, R., Piran, T. & Shemi A., Neutron star and black hole binaries in the Galaxy *Astroph. J.* **379**, L17-L20 (1991).

51 Portegies Zwart, S.F. & Yungelson, L.R., Formation and evolution of binary neutron stars *Astron. Astrophys.* **332**, 173-188 (1998).

52 Belczynski, K. , Kalogera, V. & Bulik, T., A comprehensive study of binary compact objects as gravitational wave sources: evolutionary channels, rates, and physical properties *Astrophys. J.* **572**, 407-431 (2002).

53 Belczynski, K., *et al*., Cyg X-3: A galactic double black hole or black-hole-neutron-star progenitor *Astrophys. J.* **764**, 96-102 (2013).

54 Lesh, J. R., The kinematics of the Gould Belt: an expanding Group? *Astroph. J.* **17** *(Suppl.)*, 371-444 (1968).

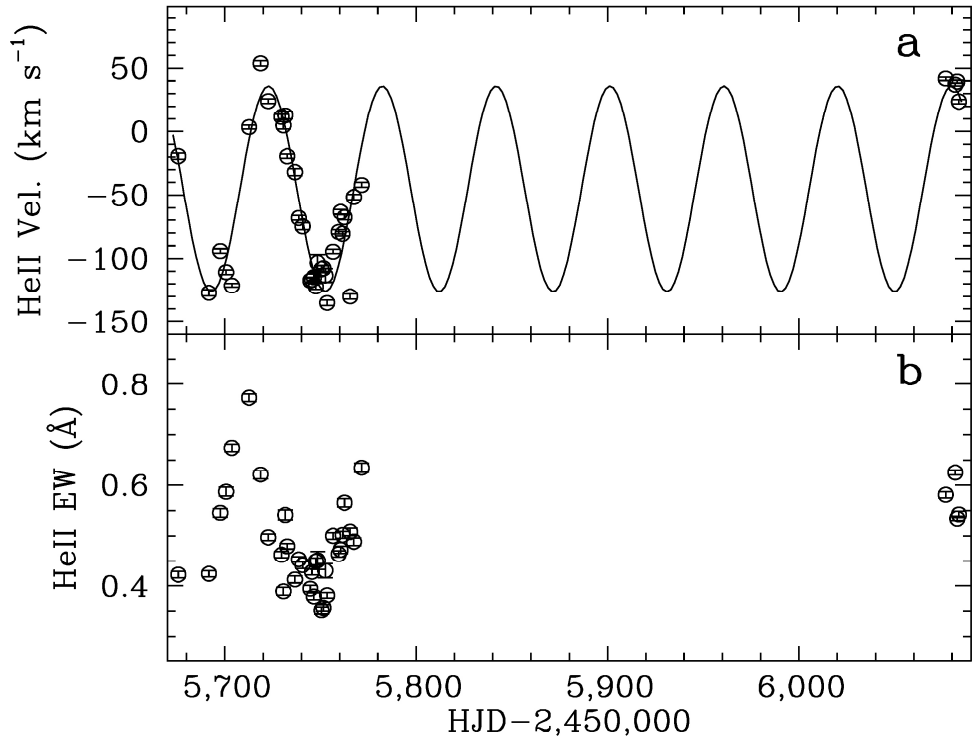

**Extended Data Figure 1: Time evolution of the He II 4,686 Å emission line in MWC 656. a,** Radial velocities obtained from single Gaussian fits to the line profile. The best fitting sine wave, with a period of 59.5 days, is overplotted. Maximum velocity occurs at HJD 2455722.2 or photometric phase 0.06. **b,** Equivalent width (EW) as a function of time. We used the convention of positive equivalent widths for emission lines. Error bars, 1 s.d.

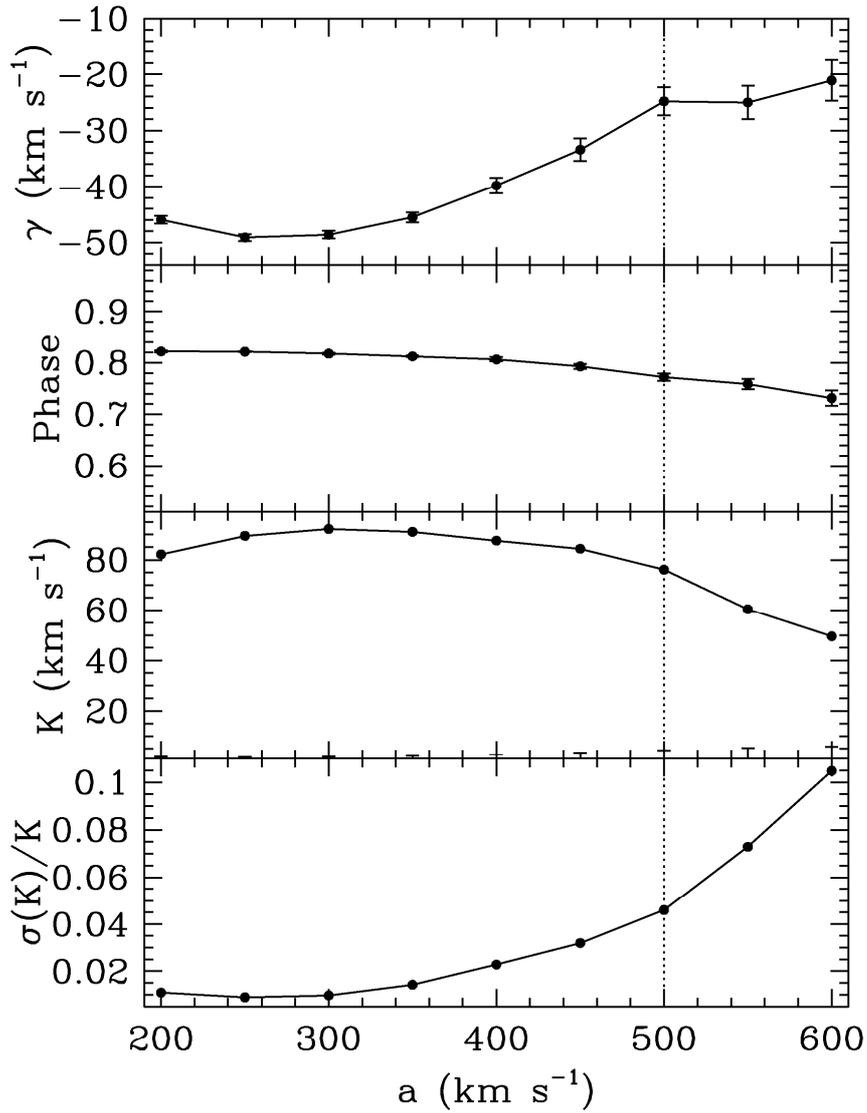

**Extended Data Figure 2: Diagnostic diagram for the He II 4,686 Å line in MWC 656.** It has been computed using the double-Gaussian technique with a Gaussian width equal to the instrumental resolution full-width at half-maximum, 55 kms$^{-1}$. The vertical dotted line indicates the Gaussian separation for which the continuum noise starts to dominate. Error bars,1 s.d. Panels display the evolution of the sinewave fitting parameters with Gaussian separation $a$. From top to bottom: the systemic velocity $\gamma$, the sinewaver phase, the velocity semiamplitude $K$ and the control parameter $\sigma(K)/K$.

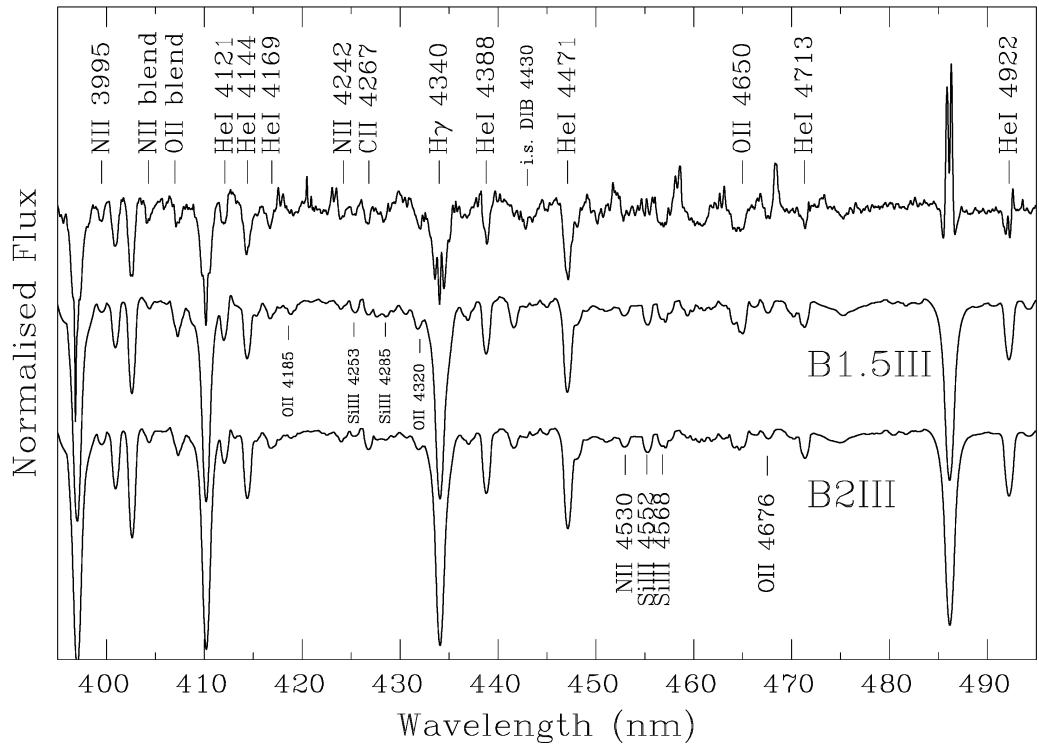

**Extended Data Figure 3: Classification spectrum of the Be star in MWC 656.** From top to bottom, spectra of MWC 656 and the MK standards HD 214993 (B1.5 III) and HD 35468 (B2 III) (ref. 54). The standards have been artificially broadened by 330 kms$^{-1}$ to mimic the rotational broadening of MWC 656.

**Extended Data Table 1: Observing log of MWC 656.**

| Date | Telescope+ instrument | Spect. Range Å | # Exp. | Integration (sec.) | Res. $\Delta\lambda/\lambda$ |
|---|---|---|---|---|---|
| 23 Apr − 28 Jul 2011 | LT+FRODOspec | 3900−5215 | 32 | 600 | 5000 |
| 28 May 2012 | LT+FRODOspec | 3900−5215 | 1 | 600 | 5000 |
| 2 − 4 June 2012 | LT+FRODOspec | 3900−5215 | 3 | 600 | 5000 |
| 26 Oct 2012 | MT+HERMES | 3770−9000 | 1 | 900 | 85000 |

**Extended Data Table 2: Orbital elements derived from radial velocities of the He II 4,686 Å and Fe II 4,583 Å lines.**

| Parameter | He II $\lambda4686$ | Fe II $\lambda4583$ |
|---|---|---|
| $P_{\mathrm{orb}}$ (days) | 60.37 (fixed) | 60.37 (fixed) |
| $T_0$ (HJD−2,450,000) | 3245.3±7.5 | 3243.1±3.7 |
| $e$ | 0.08±0.06 | 0.24±0.08 |
| $\omega$ (deg) | 351.7±44.4 | 164.4±22.1 |
| $\gamma$ (km s$^{-1}$) | −44.5±3.4 | −13.5±1.8 |
| $K$ (km s$^{-1}$) | 78.8±4.6 | 31.0±2.4 |
| $a\sin i$ (R$_\odot$) | 93.7±5.4 | 35.9±2.8 |
| $f(M)$ (M$_\odot$) | 3.02±0.53 | 0.17±0.04 |
| $\sigma_{\mathrm{f}}$ (km s$^{-1}$) | 19.2 | 10.3 |

*$T_0$ is the epoch of periastron, $e$ the orbit eccentricity, $\omega$ the longitude of periastron, $\gamma$ the systemic velocity, $K$ the velocity semiamplitude, $a$ the semimajor axis, $i$ the binary inclination, $f(M)$ the mass function and $\sigma_{\mathrm{f}}$ the r.m.s. of the fit.*